%% file: main.tex
\RequirePackage{amsmath}
\documentclass[runningheads,a4paper]{llncs}
\usepackage[title]{appendix}

\makeatletter
\newcommand{\@chapapp}{\relax}%
\makeatother

\usepackage{cmap}
\usepackage[T1]{fontenc}

\usepackage{graphicx}
\usepackage{subfigure}

\usepackage{amsfonts}
\usepackage[english]{babel}
\usepackage[%
rm={oldstyle=false,proportional=true},%
sf={oldstyle=false,proportional=true},%
tt={oldstyle=false,proportional=true,variable=true},%
qt=false%
]{cfr-lm}
\usepackage{cite}

\usepackage{paralist}

\usepackage{csquotes}

\usepackage{microtype}

\usepackage{url}
\makeatletter
\g@addto@macro{\UrlBreaks}{\UrlOrds}
\makeatother
\usepackage{diagbox}

\usepackage{xcolor}

\usepackage{hyperref}

\hypersetup{hidelinks,
   colorlinks=true,
   allcolors=black,
   pdfstartview=Fit,
   breaklinks=true}
\usepackage[all]{hypcap}

\usepackage[capitalise,nameinlink]{cleveref}
\crefname{section}{Sect.}{Sect.}
\Crefname{section}{Section}{Sections}

\usepackage{xspace}
\usepackage{mathtools}

\usepackage{lstautogobble}
\usepackage{listings}
\usepackage{color}
\usepackage{amsthm}
\usepackage{amsmath,amssymb}
\usepackage{tikz,tikz-timing}
\usepackage{xparse}
\usepackage{dsfont}
\usepackage{pgfplots}
\usepackage{stmaryrd}
\usepackage{algorithmic}
\usepackage{textcomp}
\usepackage[clock]{ifsym}
\usepackage{dsfont}
\usepackage[ruled,linesnumbered,noend]{algorithm2e}
\usepackage[capitalise,nameinlink]{cleveref}
\usepackage[inline]{enumitem}
\usepackage{comment}
\usepackage{multirow}
\pgfplotsset{compat=1.14}
\usetikzlibrary{shapes,arrows,fit,calc,positioning,decorations.pathreplacing,patterns}
\include{aux/commands}

\include{aux/style}

\newcommand{\repeatthanks}{\textsuperscript{\thefootnote}}

\begin{document}

\title{Real-time Stream-based Monitoring}
%If Title is too long, use \titlerunning
%\titlerunning{Short Title}

\author{Peter Faymonville\thanks{This work was partially supported by the European Research Council (ERC) Grant OSARES (No. 683300).} \and Bernd Finkbeiner\repeatthanks \and\\ Maximilian Schwenger\repeatthanks \and Hazem Torfah\repeatthanks}
\authorrunning{P. Faymonville, B. Finkbeiner, M. Schwenger, H. Torfah}
\institute{Saarland University \\
\email{\{faymonville, finkbeiner, schwenger, torfah\}@react.uni-saarland.de}
}
      
\maketitle

\begin{abstract}
We introduce \rtlola, a new stream-based specification language  for the
description of real-time properties of reactive systems.  The key feature is the integration of sliding windows over real-time intervals with aggregation functions into the language. Using sliding windows we can detach fixed-rate output streams from the varying rate input streams.
We provide an efficient evaluation algorithm of the sliding windows by partitioning the windows into intervals  according to a given monitor frequency. For useful aggregation functions, the intervals allow a more efficient way to compute the aggregation value by dynamically reusing interval summaries. 
In general, the number of input values within a single window instance can grow arbitrarily large disallowing any guarantees on the expected memory consumption. Assuming a fixed monitor output rate, we can provide memory guarantees which can be computed a-priori.
Additionally, for specifications using certain classes of aggregation functions, we can perform a more precise, better memory  analysis. We demonstrate the applicability of the new language on practical examples.
\end{abstract}

\section{Introduction}\label{sec:intro}
  \input{aux/introduction}

\subsection{Related Work}
  \input{aux/relatedwork}

\section{\rtlola}\label{sec:rtlola}

\input{aux/rtlola}

\section{Memory Analysis}\label{sec:memory:analysis}
  \input{aux/memory}

\section{Experiments}
  \input{aux/experiments}

\section{Conclusion}
  \input{aux/conclusion}

%%%%%%%%%%%%%%%%%%%%%%%%%%%%%%%%%%%%%%%%%%%%%%%%%%%%%%%%%%%%%%%%%%%%%%%%%%%%%%%
\newpage
\bibliographystyle{splncs03}
\bibliography{lit}

%%%%%%%%%%%%%%%%%%%%%%%%%%%%%%%%%%%%%%%%%%%%%%%%%%%%%%%%%%%%%%%%%%%%%%%%%%%%%%%

\end{document}

%% file: aux/commands.tex
\newcommand{\commentout}[1]{}

\newcommand{\rtlola}{{\sc{RTLola}}\@\xspace}
\newcommand{\lola}{{\sc{Lola}}\@\xspace}

\DeclareFontFamily{U}{MnSymbolC}{}
\DeclareSymbolFont{MnSyC}{U}{MnSymbolC}{m}{n}
\DeclareFontShape{U}{MnSymbolC}{m}{n}{
    <-6>  MnSymbolC5
   <6-7>  MnSymbolC6
   <7-8>  MnSymbolC7
   <8-9>  MnSymbolC8
   <9-10> MnSymbolC9
  <10-12> MnSymbolC10
  <12->   MnSymbolC12%
}{}
\DeclareMathSymbol{\powerset}{\mathord}{MnSyC}{180}

\newcommand\inlinelola[1]{\lstinline{#1}}
\newcommand\streamname[1]{\lstinline{#1}}

\newcommand{\LANGNAME}{Lola\xspace}

\newcommand{\ie}{i.e.\@\xspace}

\makeatletter
\newcommand{\etc}{%
    \@ifnextchar{.}%
        {etc}%
        {etc.\@\xspace}%
}
\makeatother

\newcommand{\true}{\mathit{true}}

\newcommand{\ter}{\mathit{ter}}
\newcommand{\inv}{\mathit{inv}}
\newcommand{\ext}{\mathit{ext}}

\definecolor{CommentColor}{rgb}{0.16,0.60,0.16} %red
\definecolor{eclipseBlue}{RGB}{42,0.0,255}
\definecolor{eclipseGreen}{RGB}{63,127,95}
\definecolor{eclipsePurple}{RGB}{127,0,85}

\crefname{algocf}{alg.}{algs.}
\Crefname{algocf}{Algorithm}{Algorithms}

\lstdefinelanguage{Lola}{
  keywords=[0]{constant, input, output, trigger, timeinput, frequency},
  keywordstyle=[0]\color{eclipseBlue}\bfseries,
  keywords=[1]{int, double, bool, string, ipv4, ipv6},
  keywordstyle=[1]\color{eclipseGreen},
  keywords=[2]{if, then, else},
  keywordstyle=[2]\color{eclipseBlue},
  keywords=[3]{invoke, terminate, extend},
  keywordstyle=[3]\color{eclipsePurple},
    sensitive=false,
    comment=[l]{//},
    morecomment=[s]{/*}{*/},
    morestring=[b]',
    morestring=[b]"
}

\newcommand*\pluseq{\mathrel{+}=}

\newcommand*{\Set}[1]{\{#1\}}

\newcommand{\hz}{\textbf{Hz}\xspace}

%% file: aux/style.tex
\definecolor{CommentColor}{rgb}{0.16,0.60,0.16} %red
\definecolor{eclipseBlue}{RGB}{42,0.0,255}
\definecolor{eclipseGreen}{RGB}{63,127,95}
\definecolor{eclipsePurple}{RGB}{127,0,85}
 
\lstdefinelanguage{Lola}
{
  morekeywords={
    output, if, then, else, input, trigger, const, :=, inv, ter, ext,
    int, bool, ipv4, ipv6, double, string, ephemeral, time
  },
  sensitive=true, % keywords are not case-sensitive
  morecomment=[l]{//}, % l is for line comment
  morecomment=[s]{/*}{*/}, % s is for start and end delimiter
  morestring=[b]{"} % defines that strings are enclosed in double quotes
}

%% file: aux/introduction.tex
In stream-based monitoring, we translate input streams, which contain sensor
readings and other data collected at runtime, into output streams, which
contain aggregate statistics, such as an average value, a counter, or an integral. Trigger specifications define thresholds and other logical conditions on the
values on these output streams, and raise an alarm or execute some other predefined action if the condition becomes true.
The advantage of this setup is great expressiveness and easy-to-reuse, compositional specifications.

Existing stream-based languages like Lola~\cite{Angelo+others/05/Lola,Faymonville2016} are based on the synchronous programming paradigm, where all streams are synchronized via a global clock. In each step, the new values of all output streams are computed in terms of the values of the other streams at the previous time step or at some other time step identified by an offset given as a natural number. This paradigm provides a simple and natural evaluation model that fits well with typical implementations on synchronous hardware.

In real-time applications, however, the assumption that all data arrives synchronously is often simply not true. Consider, for example, an online store like Amazon, where we might monitor the customer reviews on different products. Depending on the popularity of the product, new reviews might arrive at rates ranging from seconds to days, months, or even years. In fact, the arrival rates for a single
product would typically vary wildly, depending, for example, on the time of day.
In principle, one could, of course, run all streams at the rate of the fastest input stream. But such an approach would not only be extremely wasteful in terms of memory and processing time, it would also produce its results much too rapidly. For example, an online store might only wish to publish updated review statistics on a daily basis, not at the rate of individual incoming reviews.

In this paper, we present \rtlola, the extension of the stream-based runtime verification language Lola with real-time events. We introduce the following new concepts into Lola:

\begin{enumerate}
\item Variable-rate input streams: we consider input streams that extend at a-priori unknown rates. The only assumption is that each new event has a real-valued time stamp and that the events arrive in-order.
\item Real-time offsets: since the streams are no longer synchronous,
it is no longer sensible to refer to events in other streams via an offset given as a number of steps. Real-time offsets are given in units of time, such as 1 hour.  
\item Sliding windows: A sliding window aggregates data over a real-time window given in units of time. For example, we might count the number of reviews over the past 24 hours.
\end{enumerate}

The following example illustrates in a simple real-time controller scenario how \rtlola\ translates variable-rate input streams into a fixed-rate output streams:

\begin{lstlisting}
  input  double sensor
  input  double reference
  input  double timestamp
  output double error := sensor[2sec, 0, avg] 
                    - reference[2sec, 0, avg]
  output double acc_error := error[10sec, 0, $\int$]      
  trigger acc_error > 5
\end{lstlisting}

The output stream {\tt error} computes the average value of a sensor reading over a 2 second window and then subtracts from this the average value of a reference signal. The output stream {\tt acc\_error} then integrates over the error in a 10 second window. A trigger is raised if {\tt acc\_error} becomes too large.

The real-time features of \rtlola\ integrate seamlessly into the other specification mechanisms of Lola, such as parameterization. 
%Consider the online store described in the following example:
%
%\begin{lstlisting}
% input int prod_id, user_age
% input bool expired
% output int views<id:int>
%        inv: prod_id
%        ext: prod_id = id & user_age <30
%        ter: prod_id = id & expired
%        := 1
% output int views_per_h<id:int> 
%        inv: prod_id
%        ter: prod_id = id & expired
%        := views(id)[1h,0,sum]
% trigger any(views_per_h>100)
%\end{lstlisting}
%
%The specification defines a monitor that checks whether a product has been viewed more than 100 times by customers under the age of 30 in one hour time, and before the product expires. The monitor becomes its data from three input streams \emph{prod\_id, user\_age} and \emph{expired}. Every time a new product id $i$ is observed a new instance $\textit{views}(i)$ and $\textit{views\_per\_h}(i)$ are instantiated. The instance $\textit{views}(i)$ marks whether the product $i$ has been viewed by a customer under 30. The instance $\textit{views\_per\_h}(i)$ computes the number of views in the last hour and is extended every time  $\textit{views}(i)$ is extended. If the number of views increases over 100 in some point of time, the monitor alerts the owner of the shop. If the products expires without reaching 100 views, then the instance is terminated, and eliminated from the set of instances of \emph{views}. The rate of \emph{views\_per\_h} is determined by the extension rate of \emph{views}.

As with any semantic extension, the challenge in the design of \rtlola\ is to maintain the efficiency of the monitoring. The \emph{efficiently monitorable fragment}~\cite{Angelo+others/05/Lola} of Lola consists of all specifications that only have bounded references into the future. This fragment can be monitored with constant memory.
Obviously, not all \rtlola\ specifications can be monitored with constant memory, even if we restrict offsets to refer to the past: Since the rates of the input streams are unknown, an arbitrary number of events may occur in the span of a fixed real-time unit. We can, nevertheless, again identify an efficiently monitorable fragment that covers many specifications of practical interest.

For the space-efficient aggregation over real-time sliding windows, we partition the real-time axis into equally-sized intervals. The size of the intervals is dictated by the rate of the output streams. For certain common types of aggregations, such as the sum or the number of entries, the values within each interval can be \emph{pre-aggregated} and then only stored in this summarized form.

In a static analysis of the specification, we identify parts of the specification with unbounded memory consumption, and compute bounds for all other parts of the specification. In this way, we can determine early, whether a particular specification can be executed on a system with limited memory.

%% file: aux/relatedwork.tex
\rtlola is based on it synchronous predecessor
\lola~\cite{oldlola,lola}, where all streams are synchronized
via a global clock and the computation is performed in a single, centralized monitor, corresponding to its main application domain: monitor circuit implementations on
synchronous hardware.

%In real-time applications, however, the assumption that all data arrives synchronously is often simply not true. For example, in a communication over a data-bus, where a master requests data from several connected sensors, the distribution of addressed slaves can vary greatly depending on the master's circumstances.

%Although it
%would in principle be possible to write monitors in any
%programming language, monitors are -- to alleviate correctness and resource concerns -- often synthesized from
%a formal specification language like a temporal logic 
%%A vast number
%%of real-time logics have been proposed for specifying monitors for
%%real-time applications 
%\cite{113764, Koymans1990, Raskin1997,
%DBLP:conf/lics/HarelLP90, Donze:2010:RST:1885174.1885183,6313045}.
%The advantage of using a temporal logic is that the derived monitor is
%usually guaranteed to run with predetermined bounds on time and memory. However, most
%monitors generated from temporal logics are limited to Boolean verdicts.

%\rtlola combines the expressiveness of a programming language with the strong guarantees on the resource usage of a logic.

In the area of monitor specification languages, a series of works have investigated the monitoring of real-time properties. Many of these approaches are based on real-time variants of temporal logics \cite{113764, Koymans1990, Raskin1997, DBLP:conf/lics/HarelLP90,  Donze:2010:RST:1885174.1885183,6313045}.  Maler and Nickovic present a monitoring algorithm for properties written in signal temporal logic (STL) by reducing STL formulas via a boolean abstraction to formulas in the real-time logic MITL \cite{Maler2004}. Building on the latter ideas, Donze et. al. present an algorithm for the monitoring of STL properties over continuous signals \cite{Donz2013}. The algorithm computes the robustness degree in which a piecewise-continuous signal satisfies or violates an STL formula. Basin~et.~al. extend metric logics with parameterization \cite{DBLP:journals/jacm/BasinKMZ15}. A monitoring algorithm for the extension was implemented in the  tool MonPoly~\cite{Basin2012}. MonPoly was introduced as a tool for monitoring usage-control policies. Another extension to metric dynamic logic was implemented in the tool Aerial~\cite{aerial}.  However, most monitors generated from temporal logics are limited to Boolean verdicts.

Within stream-based monitoring approaches, two frameworks with expressive verdicts have been introduced in the \emph{Copilot} framework \cite{Pike2010} and with the tool BeepBeep 3~\cite{Hall2016}. Copilot is a stream-based dataflow language based on several declarative stream processing languages \cite{oldlola,lustre}. From a specification in Copilot, constant space and constant time C programs implementing embedded monitors are generated. 
The BeepBeep 3 tool uses an SQL-like language that is defined over streams of events. In addition to stream-processing, it contains operators such as slicing, where inputs can be separated into several different traces, and windowing where aggregations over a sliding window can be computed. However, both approaches assume a synchronous computation model, where all events arrive at a fixed rate and thus are limited to applications that have such an underlying model. 
A non-synchronized real-time stream specification language was introduced with TeSSLa \cite{DBLP:conf/sac/LeuckerSS0S18}. TeSSLa allows for monitoring piece-wise constant signals where streams can emit events at different speeds with arbitrary latencies. In contrast to \rtlola, TeSSLa does not support parametrization and sliding windows.
A similar decoupling from variable input event rates via sliding windows and fixed rate clock streams has been raised in \cite{Basin2017} within the context of temporal logics. Many works have considered approaches for the efficient evaluation of aggregations on sliding windows \cite{efficienteval}. Basin et. al. present an algorithm for combining the elements of subsequences of a sliding window with an associative operator, which reuses the results of the subsequences in the evaluation of the next window \cite{Basin:2015:GCA:2941009.2941356}.

On the application side, multi-robot systems with central task servers are a common paradigm, as shown in \cite{multi_robot_1,multi_robot_2,multi_robot_3,antlab}. Bandwidth restrictions as in our application  have also been studied in several approaches for distributed wireless sensor systems \cite{DBLP:journals/ejasp/WangEG03,collecting}.

%\rtlola extends the language Lola~\cite{Faymonville2016} to include real-time properties and subsumes  temporal logics and all languages mentioned above. It provides natural encodings for both temporal correctness properties and statistical measures. As for Lola, \rtlola has the advantage of allowing each stream to run on an individual slice of the incoming data and on an individual time scale.

%Using sliding windows might result in possibly high memory requirements due to the need to cache an unbounded number of input values.  Using these results, they show how properties of aggregated data can be monitored efficiently for an extension of metric temporal logic with aggregation operators \cite{Basin2015}. We extend this idea with a combination of fixed rate output streams and invertible aggregation functions, which allow us to summarize a variable number of input values into a single value and lifts the window aggregation computation to the interval summaries. The granularity of the intervals is inferred from the fixed output evaluation rate.

%% file: aux/rtlola.tex
\begin{figure}[h]
\centering
\begin{tikzpicture}[xscale=1,transform shape,module/.style={
draw,
fill=white,
rectangle, 
minimum width=1cm,
text width=1.3cm,
minimum height=1.5cm,
align=center,
font=\scriptsize
}]

%% central station %%
\newcommand\lx{0.2} % left x
\newcommand\rx{7.55} % right x
\newcommand\uy{0} % upper y
\newcommand\ly{-2.5} % lower y
\draw [dashed] (\lx,\uy) -- (\rx,\uy) -- (\rx,\ly) -- (\lx,\ly) -- (\lx,\uy);

\node (control) at (1.2,-1.25) [module, minimum width=1.6cm, minimum height=.75cm] {Controller};

\newcommand\mlx{3} % left x
\newcommand\mrx{7.4} % right x
\newcommand\muy{-0.1} % upper y
\newcommand\mly{-2.4} % lower y
\draw [] (\mlx,\muy) -- (\mrx,\muy) -- (\mrx,\mly) -- (\mlx,\mly) -- (\mlx,\muy);

\node (windows) at (4.0,-1.3) [module, minimum width=1.5cm, minimum height=1.2cm, double] {Sliding\\Windows};

\node (outputs) at (6.3,-1.3) [module, minimum height=1.2cm] {Outputs + Triggers};

\node (clock) at (6.3,-.34) [circle, scale=.8] {$\VarTaschenuhr$};
\draw [->] (6.3,-.52) -- (outputs);

% controller -> windows
\foreach \y in {-1, -1.25, -1.5}
  \draw [->] (2,\y) -- (3.25,\y);

% Windows -> Outputs + Triggers
\foreach \y in {-1, -1.25, -1.5}
  \draw [->] (4.79,\y) -- (5.5,\y);

% Outputs + Trigger -> Windows
\draw [->] (7.05,-1.7) -- (7.2,-1.7) -- (7.2,-2.2) -- (3.1,-2.2) -- (3.1,-1.7) -- (3.25,-1.7);

\node [] (centrstat) at (3.8,0.2) {\tiny  Central Station};

%% end central station %%

%% robots %%

%% robot 1 

%\newcommand\rolx{0.6} % left x
%\newcommand\rorx{2.6} % right x
%\newcommand\rouy{-3.4} % upper y
%\newcommand\roly{-5} % lower y
%\draw [dashed] (\rolx,\rouy) -- (\rorx,\rouy) -- (\rorx,\roly) -- (\rolx,\roly) -- (\rolx,\rouy);
%
%\node [draw] (c1) at (1.6,-4.7) {\tiny  Controller};
%\draw [->] (1.6,-4.5) -- (1.6,-4.4);
%\node [draw] (w1) at (1.6,-4.2) {\tiny  Sliding Windows};
%\draw [->] (1.6,-4.0) -- (1.6,-3.9);
%\node [draw] (o1) at (1.6,-3.7) {\tiny  Outputs + Triggers};
%\draw [->] (o1) -- (windows);
%
%\node [] (r1) at (1.6,-5.2) {\tiny  Robot 1};
%
%\node (dots) at (3.8,-4.2) {...};
%
%%% robot 2
%\newcommand\ronlx{5} % left x
%\newcommand\ronrx{7} % right x
%\newcommand\ronuy{-3.4} % upper y
%\newcommand\ronly{-5} % lower y
%\draw [dashed] (\ronlx,\ronuy) -- (\ronrx,\ronuy) -- (\ronrx,\ronly) -- (\ronlx,\ronly) -- (\ronlx,\ronuy);

%\node [draw] (cn) at (6,-4.7) {\tiny  Controller};
%\draw [->] (6,-4.5) -- (6,-4.4);
%\node [draw] (wn) at (6,-4.2) {\tiny  Sliding Windows};
%\draw [->] (6,-4.0) -- (6,-3.9);
%\node [draw] (on) at (6,-3.7) {\tiny  Outputs + Triggers};
%\draw [->] (on) -- (windows);

%\node [] (r2) at (6,-5.2) {\tiny  Robot $n$};
%% end robots %%

\end{tikzpicture}
  
\caption{The \rtlola framework}
\label{fig:framework} 
\end{figure}  

\rtlola provides the stream-based monitoring framework depicted in Figure~\ref{fig:framework}. In stream-based monitoring, sensor readings and other collected data at runtime are interpreted as streams. These \emph{input streams}, are translated to new \emph{output streams}, which define filters and aggregated statistics, such as an average value, a counter, or an integral. Monitoring tasks are defined by so-called \emph{trigger} specifications, which define thresholds and other logical conditions on the values of input and output streams. The advantage of this setup is great expressiveness and easy-to-reuse, compositional specifications. Monitoring an \rtlola specification can be distributed into monitoring tasks that are executed on a central server we call the \emph{central monitor}. The central monitor evaluates data received from the central controller of the multi-robot system and from data collected from the robots themselves. Data collected from the robots can in turn be processed before it is sent to the central monitor. This is done by applying parts of the \rtlola specifications locally on the robots. In general the goal is to relief the network from intensive data transmission. The \rtlola framework provides tool support to determine the cost of executing monitoring tasks. 

In this section, we introduce the specification language of the \rtlola framework and provide an evaluation algorithm for executing the monitoring tasks. In the next section we show how we can implement an automatic memory analysis procedure on top of the language that analyzes the memory consumption of an \rtlola specification.

\subsection{\rtlola: The Language}

\rtlola (short for Real-Time \lola) is based on the synchronous stream processing language \lola~\cite{oldlola,lola}.  \lola has been used to monitor industrial hardware systems, such as bus protocols and memory controllers, as well as to monitor networks. \rtlola extends \lola \cite{lola} with two new key features that make it possible to monitor real-time requirements: \emph{sliding window} expressions, which compute aggregated values over windows of fixed (real-time) durations, and \emph{stream computation rates}, which define the pace in which streams are extended with new values.

An \rtlola specification is a system
of stream declarations over typed \emph{stream variables}. This system consists
of $m$ typed input stream declarations of the form:
\begin{lstlisting}
  input $\ T_i\ t_i $
\end{lstlisting} 
where each stream $t_i$ is of type $T_i$ for $i \leq m$, and $n$ typed parameterized output stream templates of the form 
\begin{lstlisting}
  output $T_{m+j}$ $s_j$ $\langle T^1_j p_1,\dots,T^{k_j}_j p_{k_j} \rangle : C$
    invoke: $\mathit{inv}_j$
    extend: $\mathit{ext}_j $
    terminate: $\mathit{ter}_j$
    := $e_j(t_1, \dots, t_m, s_1, \dots, s_n, p_1, \dots p_{k_j})$
\end{lstlisting}
for $j \leq n$. Each stream template introduces a \emph{template variable} $s_{j}$ of type $T_{m+j}$ that depends on \emph{parameters} $p_1,\dots,p_{k_j}$ of types $T_{j}^{1}, \dots T_{j}^{k_j}$, respectively.
For given values $\alpha = (v_1,\dots,v_{k_j})$ of matching types $T_{j}^{1}, \dots T_{j}^{k_j}$, we call a stream $s_j(\alpha)$ an \emph{instance} of the template $s_j$ if it computes the stream induced by 
$ e_j(t_1,\dots,t_m, s_1, \dots s_n, v_1, \dots, v_{k_j})$. The stream $s_j(\alpha)$ is created, extended, and deleted according to the following \emph{auxiliary streams}:
  \begin{itemize}
    \item $\mathit{inv}_j$ is the \emph{invocation} template stream variable of $s_j$ and has type $T_j^{1}\times \dots \times T_j^{k_j}$. If an instance of $\mathit{inv}_j$ is extended with a fresh value $(v_1, \dots v_{k_j})$, then an instance $s_j({v_1,\dots,v_{k_j}})$ of $s_j$ is invoked unless it already exists.
    \item $\mathit{ext}_j$ is the \emph{extension} template stream variable of $s_j$ and has type \emph{bool}. If an instance $s_j(\alpha)$ of $s_j$ is invoked with parameter values $\alpha = (v_1,...,v_{k_j})$, then an extension stream $\mathit{ext}_j(\alpha)$ is invoked with the same parameter values~$\alpha$. The instance $s_j(\alpha)$ is extended whenever $\mathit{ext}_j(\alpha)$ is extended with \emph{true}.
    \item $\mathit{ter}_j$ is the \emph{termination} template stream variable of $s_j$ and has type \emph{bool}. If $s_j$ is invoked with parameter values $\alpha = (v_1,...,v_{k_j})$, then a terminate stream $\mathit{ter}_j(\alpha)$ is invoked. If $\mathit{ter}_j(\alpha)$ is extended with \emph{true}, the output stream $s_j(\alpha)$ is terminated.  
  \end{itemize}
 
Each instance $s_j(\alpha)$ runs on a local clock paced by its corresponding extend stream $\mathit{ext}_j(\alpha)$ and by an additional optional clock $C$. The instance is extended with every tick of $C$ if and only if $\mathit{ext}_j(\alpha)$ is true. If $C$ is not given, the local instance clock always progresses equally fast as the fastest clock amongst all local clocks of the streams appearing the the expression $e_j$. 
If neither $C$ nor $\mathit{ext}_j(\alpha)$ is given the local clock in this case progresses in a discrete fashion, however, not with equidistant ticks on the real-time axis. Whenever a new value arrives for one of the sub-streams, a new tick takes place that extends $s_j(\alpha)$. Therefore, the distance in time between two consecutive ticks can be arbitrarily large or small.

Auxiliary stream expressions can be omitted if they are inferable from the context.

The stream expression $e_j(t_1, \dots, t_m, s_1, \dots s_n,p_1,\dots,p_{k_j})$,
is defined over a set of \emph{independent} input stream variables $t_1, \dots,
t_m$, a set of \emph{dependent} output stream variables $s_1, \dots , s_n$, and a set parameter variables $p_1, \dots, p_{k_j}$. Dependent stream
variables have access to past and present values of all streams. 
An \rtlola stream expression permits basic constructs such as constants, $n$-ary function symbols, conditional expressions and \emph{stream accesses} of form $s_j(v_1,\dots,v_{k_i})[-i]$, i.e., the value of stream with an offset $i$ in the past.

\rtlola extends its predecessor with sliding window expressions of the form $s_j[r,\gamma]$, where $r$ is a real number defining the duration of the window and $\gamma$ is a function of type $(T_{m+j})^* \rightarrow T'$ defining the aggregation function. 

 Expressions of the form $e?d$ evaluate to the default value $d$ in case $e$ is not defined. This case occurs if the respective instance is not invoked at this point in time or its extend stream for an expression $s_j[-i]$ was \emph{true} less than $i+1$ times.

Finally \rtlola specifications  contain a list of \emph{triggers}:
\begin{lstlisting}
  trigger $\varphi_i$
\end{lstlisting}
where $\varphi_i$ is an expression of type boolean over stream variables. Triggers generate notifications when their expression evaluates to $\mathit{true}$. Trigger expressions also permit special operators arguing over stream instances, such as \lstinline{any} of type \lstinline{bool} checking the validity of a condition for \emph{any} existing stream instance, and \lstinline{count} of type \lstinline{int} counting the number of existing stream instances of a given stream.

As a concrete example, a parametric variant of the specification presented before can be declared as following:

\begin{lstlisting}
input bool offRoad 
input int CID
input bool pickUp

output bool offRoadPickUp<int cid>: 0.1Hz
  invoke: CID
  extend: cid = CID    
  := offRoad & pickUp

output bool suspicious<int cid> :=
  invoke: CID
  extend: cid = CID
  := offRoadPickUp(cid)[8h, count]?0 > 5
\end{lstlisting}

Observe that the output streams are now parametric in the robot car identifier \texttt{cid}. Since they have \texttt{CID} as an invoke stream, new stream instances of both output streams are created whenever a fresh identifier appears on the input stream \texttt{CID}. The extend stream then extends those stream instances whenever the matching identifier \texttt{cid} appears again on the input stream \texttt{CID}. Within the parametric output stream \texttt{suspicious<int cid>}, we access the corresponding \texttt{offRoadPickUp} stream instance by referring to \texttt{offRoadPickUp(cid)}.

\subsection{Efficient Evaluation}

For a space and time efficient evaluation, we need to pay special attention to the computation of sliding windows. Consider a sliding window over a target stream $s$ with aggregation function $\gamma$ and window size $S$. In a na\"ive approach, whenever $s$ produces a new value, the value is stored until $S$ has passed. In the meantime, when the window is evaluated, all stored values are aggregated using $\gamma$. For input streams with variable input rate, storing all values while retaining strict memory guarantees is not an option. 

We apply a paning approach~\cite{efficienteval}. Here, we split $S$ into a pre-determined amount of time frames with equal width. All values arriving within the same frame are pre-aggregated using $\gamma$. The resulting value now represents all values that arrived within the frame, enabling us store only one value rather than all incoming values. When the window is evaluated, the pre-aggregated values are aggregated once again. 
Pre-aggregation is possible for \emph{homomorphic}  functions $\gamma^*$. Most widely used aggregation functions such as \textit{max}, \textit{min}, \textit{sum}, and \textit{integral} are in fact homomorphic. 

%The drawback of using time frames is that we lose information about single values, \ie,  when evaluating a window, the entire frame is either considered or disregarded.

% Here, we could discuss left-invertibility if we'd like to.

Window paning greatly reduces worst-case memory consumption for variable rate target streams. More details can be found in \Cref{sec:memory:analysis}. Another concern for safety monitoring is the running time: If the computation takes too long, it might already be too late when the monitor detects the problem.

Although the running time for each stream expression in isolation can be bounded quite easily and is constant in the length of the trace, this is not the case of the number of stream instances due to parametrization. 

Consider for example a parametrized stream counting the number of times one specific car picked up a customer off road:

\begin{lstlisting}
  input int CID
  
  output int suspicious <int cid>
    invoke: CID
    extend: CID = cid
    := offRoadPickup[8h, count(cid), 0]
\end{lstlisting}

Here, the number of instances of \streamname{suspicious} depends on the number of different car ids in the trace, bounded by the domain of int or the number of cars in the fleet, whichever is less.

A large number of instances can have a significant impact on the runtime if each instance needs to be evaluated at the same time.

Note, however, that in this example the extend condition for the stream, \inlinelola{CID = cid}, indicates that only one instance can be extended at any time: If a new value $\mathit{nCID}$ of \streamname{CID} arrives, there can only be one instance of \streamname{suspicious} with parameter $\mathit{cid} = \mathit{nCID}$ that needs to be extended.

We say that a stream is \emph{efficiently bound} if the extend condition is a positive boolean formula where the atoms are equality checks between parameter values and input streams. Extending efficiently bound streams does not depend on the number of active instances and, thus, has constant running time.

\subsection{Evaluation Algorithm}
\label{sec:algorithm}

The evaluation algorithm consists of two components, one running periodically and computing the value of fixed-rate streams, and one running sporadically whenever new input values arrive.

\input{aux/figures/varratealgo}

\Cref{alg:varrate} outlines the variable-rate component. It is triggered whenever a new input arrives. First, the respective input streams are extended and new values are added to all depending window instances. Note that this does not trigger the computation of the overall window value, but is local to a single interval. Then, streams which invoke stream were extended are identified and invoked. Afterwards, efficient streams are extended. As discussed before, the existential quantifier can be resolved efficiently by the efficient binding. Again, depending windows are notified and respective instances invoked. Note that efficient streams cannot be invoked by other efficient streams. As a result, one iteration over $\mathit{EfficientStreams}$ suffices to cover all extended instances.

The last step identifies active triggers efficiently by only considering count- and exists-triggers whiles targets were invoked. For any-triggers only one instance per stream template and input even can be extended, rendering the check for active triggers efficient.

The running time of \Cref{alg:varrate} is very low, because 
\begin{enumerate}[label=(\roman*)]
  \item the number of invokes and extends in one execution is strictly limited,
  \item registering values in sliding windows only requires few arithmetic instructions, and
  \item the overhead induced by the loops in lines 5 and 12 can be greatly reduced by using clever book-keeping data structures.
\end{enumerate}

\input{aux/figures/fixedratealgo}

\Cref{alg:fixrate} describes the fixed-rate component. Here, the update of windows, extension, and invocation of streams is repeated until no further action is possible, \ie, a fixpoint is reached. This process keeps track of all extended streams and their values. Reaching a fixpoint triggers the deletion of all instances with an active termination condition. The last step returns all active triggers and newly computed output values.

%% file: aux/figures/varratealgo.tex
\begin{algorithm}
  \KwIn{Events $\mathit{e_i}$ for Streams $s_i \in S \subseteq \mathit{InputStreams}$, Timestamp $\mathit{ts}$}
  \KwResult{Trigger}
  
  \SetKw{InvokedBy}{InvokedBy}
  \SetKw{ExtendedBy}{ExtendedBy}
  \SetKw{Instances}{Instances}
  \SetKw{Is}{is}

  \Begin{
    \ForEach{$s_i \in S$}{
        extend $s_i$ by value $e_i$\;
        register $e(s)$ in depending windows at time $\mathit{ts}$\;
      \ForEach{$s \in \InvokedBy(s_i)$}{
          invoke $s$ with parameter $e_i$\;
%          $i$ = invoke $s$ with parameter $e_i$\;
%          \If{$i$ \Is EfficientStream}{$\mathit{EfficientInstances} \pluseq i$}
%          \Else{$\mathit{OutputInstances} \pluseq i$}
      }
    }
    \ForEach{$s \in \mathit{EfficientStreams}$}{
        \If{$\exists x \in \Instances(s)$ compatible with $e$}{
            $v$ = compute value for $x$ with $e$\;
            extend $x$ by $v$\;
            register $v$ in depending windows at time $\mathit{ts}$\;
            \ForEach{$s \in \InvokedBy(s_i)$}{
              invoke $s$ with parameter $v$\;
            }
        }
    }
    \Return{active triggers}
  }  
  \caption{\textsc{VarRateEvaluation}}
  \label{alg:varrate}
\end{algorithm}

%% file: aux/figures/fixedratealgo.tex
\begin{algorithm}
  \KwIn{Clock signal at time ts}
  \KwResult{Trigger, Outputs}
  \SetKw{InvokedBy}{InvokedBy}
  \SetKwBlock{Fixpoint}{fixpoint}{end}

  \Begin{
%    $\mathit{Extends} := \{s^\alpha \in \mathit{OutputInstances} \mid \mathit{ext}_s^\alpha[0] = \true \}$\;
%    $\mathit{Terminates} := \{s^\alpha \in \mathit{OutputInstances} \mid \mathit{ter}_s^\alpha[0] = \true \}$\;
    $\mathit{NewOutputs} := \emptyset$\;
    \Fixpoint{
        \ForEach{$w \in \mathit{Windows}$}{
           Compute latest value for $w$ at time $\mathit{ts}$
        }
%        \ForEach{$s \in$ Extends } {
        \ForEach{$s^\alpha$ with $\mathit{ext}_s^\alpha[0] = \true$} {
            $v$ = compute value for $s^\alpha$\;
            extend $s^\alpha$ by $v$\;
            register $v$ in depending windows at time $\mathit{ts}$\;
            $\mathit{NewOutputs} \pluseq (s^\alpha,v)$\;
            \ForEach{$s' \in \InvokedBy(s)$} {
                invoke $s'$ with parameter $v$\;
            } 
%            \If{$v == \true$}{
%                \ForEach{$y \in$ ExtendedBy($s$)}{
%                    $\mathit{Extends} \pluseq y$
%                }
%                \ForEach{$y \in$ TerminatedBy($s'$)}{
%                    $\mathit{Terminates} \pluseq y$
%                }
%            }
        }
    }
%    \ForEach{$s \in \mathit{Terminates}$}{
    \ForEach{$s^\alpha$ with $\mathit{ter}_s^\alpha[0] = \true$}{
        terminate $s$
    } 
    \Return{active triggers, $\mathit{NewOutputs}$}
  }  
  \caption{\textsc{FixedRateEvaluation}}
  \label{alg:fixrate}
\end{algorithm}

%% file: aux/memory.tex
\subsection{Dependency graph}
 Let $\varphi$ be a specification with input stream variables $t_1, \dots, t_m$ and template stream variables $s_1, \dots, s_n$ with invoke, extend and terminate stream $\inv_i, \ext_i, \ter_i$, and stream expression $e_i$ for $i \leq n$. 
An \emph{annotated dependency graph (ADG)} is a directed graph $\mathit{ADG}_\varphi = (V,E,\pi, \lambda)$, where $V = \{t_1,\dots,t_m,s_1\dots,s_n\}$ is a set of vertices each representing an input stream or a stream template variable,
 $E = \Set{(s_i, v) \mid v \in \Set{\inv_i, \ext_i, \ter_i} \vee e_i \mbox{ accesses } v}$ a set of edges defining the dependencies between the different streams, 
 $\pi: V \rightarrow \{\mathit{var}\} \cup \mathbb{N}$  a function that annotates each vertex with either $\mathit{var}$ denoting  stream with variable rates, or with a frequency in \hz for streams with fixed rates. We define \emph{var} to be larger than any natural number. 
 Finally, the function $\lambda :E \rightarrow (\Gamma\times \mathbb{N}) \cup \mathbb{N}$ that states whether the dependency of two stream is defined via a window with a duration in milliseconds and aggregation function $\gamma \in \Gamma$, or a discrete offset.

The annotation function $\lambda$ is derived from the syntax of the specification. To compute the annotation $\pi$ of vertices we use the following rules. Let $v\in V$:

\begin{itemize}
  \item If $v$ has a fixed rate $C$ then $\pi(v)=C$.
  \item Otherwise, if $v=t_j$for $j\leq m$ then $\pi(v) = \mathit{var}$, else if $v =s_j$ for $j\leq n$ then 
  $$ \pi(v) = \max\{ \pi(v') | (v,v') \in \mathit{E}\}~  $$
\end{itemize}

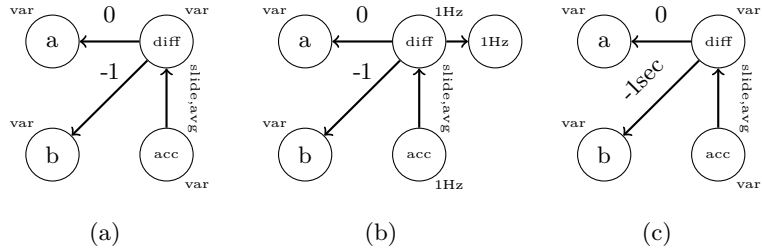
\begin{figure}[h]
\centering
	\subfigure[]{
	\begin{tikzpicture}[minimum size=0.7cm]
		\node[draw, circle](a)at(0,0){a};
		\node[](a')at(-0.4,0.4){\tiny var};
		\node[draw, circle](b)at(0,-1.5){b}; 
		\node[](b')at(-0.4,-1.1){\tiny var};
		\node[draw, circle](diff)at(1.5,0){\tiny diff};
		\node[](diff')at(1.9,0.4){\tiny var};
		\node[draw, circle](acc)at(1.5,-1.5){\tiny acc};
		\node[](acc')at(1.9,-1.9){\tiny var};
		\path[->,thick](diff) edge node [above]{0}(a);
		\path[->,thick](diff) edge node [above]{-1}(b);
		\path[->,thick](acc) edge node [above=0, rotate =-90]{\tiny slide,avg}(diff);
	\end{tikzpicture}
	}
	\subfigure[]{
	\begin{tikzpicture}[minimum size=0.7cm]
		\node[draw, circle](a)at(0,0){a};
		\node[](a')at(-0.4,0.4){\tiny var};
		\node[draw, circle](b)at(0,-1.5){b}; 
		\node[](b')at(-0.4,-1.1){\tiny var};
		\node[draw, circle](diff)at(1.5,0){\tiny diff};
		\node[](diff')at(1.9,0.4){\tiny 1Hz};
		\node[draw,circle](hz)at(2.5,0){\tiny 1Hz};
		\node[draw, circle](acc)at(1.5,-1.5){\tiny acc};
		\node[](acc')at(1.9,-1.9){\tiny 1Hz};
		\path[->,thick](diff) edge node [above]{0}(a);
		\path[->,thick](diff) edge node [above]{-1}(b);
		\path[->,thick](acc) edge node [above=0, rotate =-90]{\tiny slide,avg}(diff);
		\path[->,thick](diff)edge node {}(hz);
	\end{tikzpicture}
	}
	\subfigure[]{
	\begin{tikzpicture}[minimum size=0.7cm]
		\node[draw, circle](a)at(0,0){a};
		\node[](a')at(-0.4,0.4){\tiny var};
		\node[draw, circle](b)at(0,-1.5){b}; 
		\node[](b')at(-0.4,-1.1){\tiny var};
		\node[draw, circle](diff)at(1.5,0){\tiny diff};
		\node[](diff')at(1.9,0.4){\tiny var};
		\node[draw, circle](acc)at(1.5,-1.5){\tiny acc};
		\node[](acc')at(1.9,-1.9){\tiny var};
		\path[->,thick](diff) edge node [above]{0}(a);
		\path[->,thick](diff) edge node [above, rotate=45]{-1sec}(b);
		\path[->,thick](acc) edge node [above=0, rotate =-90]{\tiny slide,avg}(diff);
	\end{tikzpicture}
	}
\caption{Annotated Dependency Graph (ADG) to analyze the memory requirements of specifications $\varphi$, $\varphi'$, and $\varphi''$.}
\label{fig:ADG}
\end{figure}

Consider for example the following \rtlola\ specification $\varphi$
\begin{lstlisting}
	input  double a
	input  double b
	output double diff = abs(a - b[-1,0])
	output double acc = diff[10sec,avg,0] 
\end{lstlisting} 
The ADG of $\varphi$ is given in Figure~\ref{fig:ADG}(a). Both $a$ and $b$ are inputs with variable rates. Because $\mathit{diff}$ is extended every time one of the input streams is extended, its rate is also variable, and so is the one of \emph{acc}. 

Let $\varphi'$ be as $\varphi$ where \emph{diff} is supplied with a stream clock as an extension stream, then its rate becomes fixed and its expression is evaluated only once every second. The rate of \emph{acc} also becomes fixed, because it only depends on the rate of \emph{diff}. The ADG of $\varphi'$ is shown in Figure~\ref{fig:ADG}(b).

Both $\varphi$ and $\varphi'$ of the property can be computed efficiently, because the discrete offsets require only constant memory and the sliding window with the aggregation function \emph{avg} can be computed as we have seen in the last section. 

In contrast, a specification $\varphi''$, which access a variable input stream with a real-time offset operator, do not permit an a-priori memory guarantee.   
\begin{lstlisting}
	input  double a
	input  double b
	output double diff = abs(a - b[-1sec,0])
	output double acc = diff[10sec,avg,0] 
\end{lstlisting} 
Here, there is no bound on the number of events that need to be stored for computing $b[-1\mathit{sec},0]$, because it is impossible to determine beforehand which points in time will be needed in the evaluation of this expression. The ADG for $\varphi''$ is depicted in Figure~\ref{fig:ADG}(c).
Again, if we supply \emph{diff} with a stream clock, then we only have to store the values of $b$ at the coarser granularity. 
Similarly, when we access a variable rate stream with an aggregation function, which requires to store all events in a sliding window, like \emph{max}, we cannot provide a memory guarantee.

\subsection{Computing memory requirements}
We describe a concrete analysis method to derive tight memory requirements for a given \rtlola specification. Memory bounds for discrete-time offsets can be computed as for \textsc{Lola}~\cite{oldlola}. We extend this algorithm with new computational rules to determine the memory bounds for real-time sliding window expressions. For each edge $e=(v,v')$ in the ADG we can determine how many values must be stored for the computation of $v$ using the following rules, where $r$ is the duration of the sliding window:
\begin{center}%
    \begin{tabular}{|c||c|c|}
      \hline
      \diagbox[width=5em]{\small $\pi(v')$}{\small $\lambda(e)$}  & ($\gamma, r)$ & ($\gamma^*, r)$, pane size = $z$ \\
      \hline
      $\mathit{var}$\hspace{2em}&  unbounded & $\max\Set{1, rz^{-1}}$ \\
      \hline
      $\mathit{y}$Hz\hspace{2em}&  $yr$ & $\min\Set{rz^{-1}, ry}$ \\
      \hline
    \end{tabular}
\end{center}

If the aggregation function $\gamma$ requires to store all values, such as \emph{median}, and the target stream has a variable output rate, the required memory is unbounded. If the target is a fixed rate stream, and the duration of the sliding window is $r$, at most $yr$ values can arrive, thus giving a bound on the required memory.

For efficient aggregation functions $\gamma^*$ such as \emph{sum}, \emph{count}, or \emph{average}, \rtlola can use a \emph{paning} approach presented earlier. Here, the real timeline is split into intervals of equal size $z$. All values arriving within the same interval are considered equal in terms of arrival time. The discretization allows for giving tight bounds on the required memory because all values within an interval can be pre-aggregated. As a result, a single value is sufficient to represent the whole interval. Thus, even if the input rate is unbounded, $rz^{-1}$ pre-aggregated values are required. For a fixed rate input, if the rate of $v$ is sufficiently low, it is possible to store all values and forego paning. 

Lastly, let $\mu(v)$ be the values determined by the table above, and $\eta(v)$ the maximal number of instances of a stream, the upper bound for the memory consumption of a stream is $\sum_{v \in V}\mu(v)\eta(v)$.

%% file: aux/experiments.tex
We implemented \rtlola in C for unix systems. It features two modes of output
computation: In the fixed step mode, each arrival of an input value triggers a
new evaluation cycle meaning that new output values are computed. In the
variable step mode, evaluation cycles are time-triggered at a fixed frequency\footnote{The delay is subject to the non-deterministic sleep
behavior of non-real time operating systems.}. 

To realize this, we use multithreading with a \emph{reader} thread reading input
values from a .csv file.
The \emph{worker} thread is time triggered in the fixed step mode, or
event triggered by the arrival of input values in the variable step mode. We use
POSIX mutexes and and conditions for thread synchronization.

The parser for the specification is generated using the BNFC tool\footnote{
\url{http://bnfc.digitalgrammars.com/}}.

Each stream instance has a value store being either a cyclic buffer when there
are no real time accesses, so the required space is bounded. Otherwise the
store is a dynamic array of values and their respective time stamp.

Stream instances are managed in hash-maps where the parameter values are hashed
with a generic hash function.

\subsection{PID controller}

%\begin{figure}[t]
%    \input{integral_example}
%    \label{fig:example:integral}
%    \caption{Graph for sensor data derivation with accessible data points in red.}
%\end{figure}

    In the first scenario we assume a PID controller monitoring the accumulated
    error of sensor data with respect to reference values. Data points arrive
    sporadically. 
    We use the
    following specification:

    \begin{lstlisting}
input double temperature
input double reference
time input double timestamp

output double smooth_temp := temperature[10s,avg,0.0]
output double smooth_ref := reference[10s,avg,0.0]

output double error := smooth_temp - smooth_ref
output double acc_error := error[50s, avg, 0.0]

trigger any(acc_error > 0.016)
    \end{lstlisting}

    % \maximilian{TODO: Check whether this explanation is superfluous.}
%    Internally, a sound split for the integral is used. The map function $\gamma_*$
%    computes the area covered by the trapezoid induced by the latest and the
%    penultimate data value. We abbreviate \STREAM{sensor} with $s$ and
%    \STREAM{timestamp} with $t$.
%    \[ 
%        \gamma_*(s,t) = \frac{s[-1]-s[0]}{2} \cdot (t[0] - t[-1])
%    \]
%    As a result, we can use $\binop = +$ as reduction function, $\gamma_F = \mathit
%    {id}$ for no-op finalization, and $\binoplinv = -$ as left-inverse.
%
%    The computation of the window requires to store the two most recent input
%    values and the bucket values only. It is thus independent of the input
%    frequency in the asynchronous mode.
%    % \maximilian{ODOT}

    We generate input data using a Simulink feedback controller model
    % \maximilian{TODO:cite correctly} 
    and compare the fixed step mode against the variable step one on an input file with around 400 values. The results
    can be found in Table~\ref{tab:pid:results}.

    \begin{table}[ht]
    \centering
    \begin{tabular}{|c|c|c|c|c|}
        \hline
        &frequency&time[s]&memory[MB] \\
        \hline
        Variable & n/a& <1 &3.36 \\
        \hline
        \multirow{3}{*}{Fixed} & 1Hz & <1 & 3.61 \\
         &0.2Hz & <1 & 3.56   \\
         &0.1Hz & <1 & 3.55 \\
        \hline
    \end{tabular}
    \caption{Experimental results for the PID controller data.}
    \label{tab:pid:results}
    \end{table}

As the output frequency decreases, we can observe an improvement in the memory consumption of the monitor.
  
\subsection{Amazon Review Data}
    In the second scenario we consider amazon product review data, especially the review rating of products. When a significant increase in ratings for a product is
    detected, the product is considered interesting right now, so the stock of the product should be increased. 

    In the corresponding specification seen in Figure~\ref{spec:amazon}, we parametrize over each product and
    terminate instances if the arrival frequency of new reviews drops below a
    certain rate. We compute the average star rating of a product and compare the average of a long running window to the average rating in the last tenth of that window and detect a significant short-term increase. If the comparison rises above a threshold, a trigger message is
    sent.

    We again compare the fixed step and variable step mode. 
    
    We use real user data from amazon\cite{amazondata} for the category of videos. It contains a dataset with 100k elements. The results can be found
    in Table~\ref{tab:amazon:results}.

    \begin{table}[ht]
    \centering
    \begin{tabular}{|c|c|c|c|c|}
        \hline
        &frequency&time[s]&memory[MB] \\
        \hline
        Variable & n/a& 3.16 & 89.19 \\
        \hline
        \multirow{3}{*}{Fixed} & 0.004Hz & 175.2 & 1147 \\
         &0.002Hz & 204.84 & 886.1   \\
         &0.001Hz & 189.82 & 726.98  \\
        \hline
    \end{tabular}    \caption{Experimental results for the amazon review data.}
    \label{tab:amazon:results}
    \end{table}
    
    We again see that lowering the output frequency of the monitor results in better memory usage. Remember that the variable mode has no guarantees on memory consumption, even though it may be more efficient in some cases.

\begin{figure}
\begin{lstlisting}
input string uid
input string pid
input double stars
time input double timestamp

const double ter_thres := 10000.0

output bool right_pid <string prod> 
	:inv: pid
	:ter: ter_cond
	:= pid = prod

output double time_p <string prod>
	:inv: pid
	:ext: right_pid
	:ter: ter_cond
	:= timestamp

output double arr_freq <string prod>
	:inv: pid
	:ext: right_pid
	:ter: ter_cond
	:= time_p(prod)[slide, 1000, count, 0.0]

output bool ter_cond <string prod> 
	:inv: pid
	:ter: ter_cond
	:= arr_freq(prod)[0, 50000.0] < ter_thres

output double star_avg_l <string prod>
	:inv: pid
	:ext: right_pid
	:ter: ter_cond
	:= stars[slide, 100000, avg, 0.0]

output double star_avg_s <string prod>
	:inv: pid
	:ext: right_pid
	:ter: ter_cond
	:= stars[slide, 10000, avg, 0.0]

output double avg_delta <string prod>
	:inv: pid
	:ext: right_pid
	:ter: ter_cond
	:= star_avg_l(prod)[0, 0.0] - star_avg_s(prod)[0, 0.0]

trigger any(avg_delta > 1.0)
\end{lstlisting}
\caption{\LANGNAME specification for the detection of an unusual high increase
in star ratings of any product.}
\label{spec:amazon}
\end{figure}

%% file: aux/conclusion.tex
We introduced \rtlola, a stream-based specification language for the description of monitors for real-time properties.  \rtlola\ specifications allow the definition of real-time offsets into the past and the future, the aggregation over real-time sliding windows, and can handle input streams of variable rate. 
%In contrast to the discrete-time stream-based specifications, in \rtlola\ we cannot immediately provide memory guarantees. 
We showed that for many aggregation functions used in practice, we can efficiently compute the aggregations of sliding windows on variable rate input streams by partitioning the real-time axis into intervals of equal sizes, which is determined by the given output rate. 
We also showed how a static analysis can be done for a given specification to indicate to the user which parts of his specification can be computed efficiently and which might lead to high memory consumption.

This new output-driven perspective gives a pragmatic way to give a memory guarantee for the monitor and the given specification without syntactically restricting the expressiveness of the underlying stream-based specification language.

In practice, most approaches would fix a rate on the evaluation of output streams. In this way, even if the rate of input streams is variable, the evaluation of output streams remains efficient.